\documentclass[aps,preprint,groupedaddress,showpacs]{revtex4-1}

\usepackage{graphicx}
\usepackage{color}
\usepackage{comment}
\usepackage{subfig}
\usepackage{amsmath}

\begin{document}


\title{Competition Between Homophily and Information Entropy Maximization in Social Networks}


\author{Jichang Zhao$^{1,\star}$, Xiao Liang$^{2}$ and Ke Xu$^{3}$}
\affiliation{$^1$ School of Economics and Management, Beihang University \\
$^2$ Key Laboratory of Technology in Geo-spatial Information Processing and Application System, Institute of Electronics, Chinese Academy of Sciences\\
$^3$ State Key Lab of Software Development Environment, Beihang University\\
$^\star$Corresponding author: jichang@buaa.edu.cn}


\date{\today}

\begin{abstract}
In social networks, it is conventionally thought that two
individuals with more overlapped friends tend to establish a new
friendship, which could be stated as homophily breeding new
connections. While the recent hypothesis of maximum information entropy 
is presented as the possible origin of effective navigation in
small-world networks. We find there exists a competition between information entropy maximization and homophily in local structure through both theoretical
and experimental analysis. This competition means that a newly built
relationship between two individuals with more common friends would
lead to less information entropy gain for them. We conjecture that
in the evolution of the social network, both of the two assumptions
coexist. The rule of maximum information entropy produces weak ties in the
network, while the law of homophily makes the network highly
clustered locally and the individuals would obtain strong and trust ties. Our findings shed light on the social network modeling from a new perspective.

\end{abstract}


\maketitle

\section{Introduction}
\label{sec:motivation}

The last decade has witnessed tremendous research interests in
complex networks~\cite{newman_survey,evolution-zhou,xu-finacial-network}, including the
evolution of social networks~\cite{newman_cluster,jin_structure,homophily,yang_growing,clustering_citation}. It has been found that in many social networks from
different circumstances, the probability of having a friend at a
distance $r$ is $p(r)\propto r^{-1}$, which is stated as the spacial
scaling law~\cite{distance_notdead}. Recent work~\cite{max_entropy}
presents a possible origin that explains the emergence of this
scaling law with the hypothesis of maximum information entropy with energy constrains. The
authors assume that human social behavior is based on gathering
maximum information through various activities and making friends is
one of them. However, it is also found conventionally that
homophily leads to connections in social networks~\cite{homophily,transitive,jin_structure,davidsen_emergence,cnn,cnnr,toivonen_model,holme_growing,link_prediction}. Homophily is the principle that a contact between similar individuals occurs at a higher rate than
among dissimilar ones~\cite{homophily}. For instance, in social
networks, two individuals with more common friends are easier to get
connected, where the number of overlapped friends could represent the strength of homophily. Both of the above rules might drive the growth of the network in local structure simultaneously, however, to our best knowledge, little has been done to unveil the relationship between them. In this paper, we try to fill this gap from the perspective of network evolution in local structure.

\section{Theoretical Analysis}
\label{sec:ta}

\begin{figure}[h]
\centering
 \subfloat[]{\includegraphics[width=1.7in]{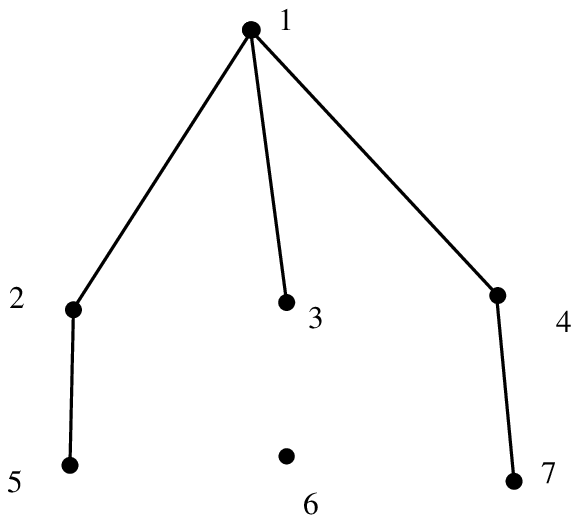}
 \label{fig:example_1}}
\subfloat[]{\includegraphics[width=1.7in]{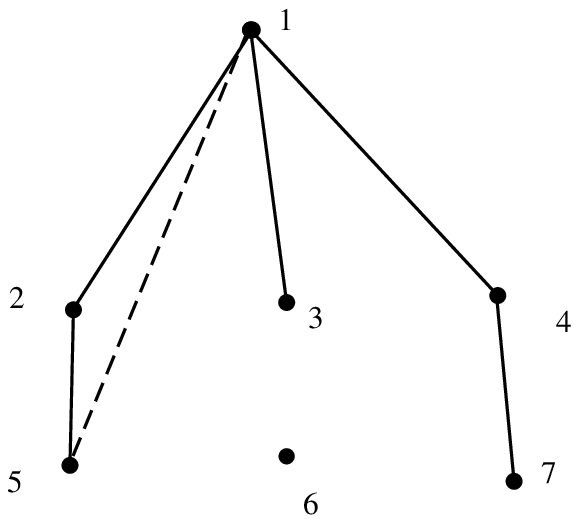}
 \label{fig:example_2}}
\caption{A simple example of the network evolution driven by homophily in local structure.\label{fig:example}}
\end{figure}

A social network can be modeled as a simple undirected graph
$G(V,E)$, where $V$ is the set of individuals (nodes) and $E$ is the
set of friendships (ties) among them. As shown in
FIG.~\ref{fig:example_1}, node 1 may obtain information from
nodes 2, 3, 4 and their friends 5, 7. Therefore, as defined
in~\cite{max_entropy}, the information sequence for node 1 is \{2,
3, 4, 5, 7\} and the frequency of each node appears in the sequence
is $q_2=q_3=q_4=q_5=q_7=1/5$ for nodes 2, 3, 4, 5 and 7
respectively, while $q_6=0$ for node 6. Then the information entropy
for node 1 can be obtained as $$\epsilon
(1)=-\sum_{i=1}^{7}{q_i\log{q_i}}=1.61.$$ Next, we assume the social
network evolves to the one as shown in FIG.~\ref{fig:example_2}
under the rule of homophily. For example, node 1 and node 5 may
establish a new friendship because they share the common friend node
2. Therefore, the updated information sequence for node 1 is \{2, 3,
4, 5, 5, 7, 2\} currently. Then the new frequency of each node
appears in the sequence is $q_2{'}=q_5{'}=2/7$,
$q_3{'}=q_4{'}=q_7{'}=1/7$, and $q_6{'}=0$. We recompute the
information entropy of node 1 as depicted above and obtain
$$\epsilon^{'}(1)=-\sum_{1}^{7}{q_{i}{'}\log{q_{i}{'}}}=1.55.$$ It
can be easily observed that $\Delta
\epsilon(1)=\epsilon{'}(1)-\epsilon(1)<0$ after node 1 built a new
tie with node 5, which means in the evolution dominated by homophily,
the information entropy for node 1 decreases. It is an intuitive
observation that the rule of homophily is incompatible with the law
of maximum information entropy, and a general explanation is introduced as
follows. Note that here we mainly discuss the network evolution in local structure, in which ties are newly built only with nodes two hops away. Because of this, with the aim of simplification, conditions of limited energy and nodes' distances are not considered in the following analytical framework. Besides, the magnificent development of the online social network has facilitated our daily social activity greatly\cite{trustties,ahn-cyworld-wwww}, so here the cost of establishing a new tie is assumed to be a constant and it is independent to the distance in social networks.

We define $n(i)$ as the set of individual $i$'s initial friends and
$k_i$ is $i$'s degree, i.e., the number of its friends. Then the set of overlapped friends between $i$ and $j$ is $c(i,j)=n(i)\cap n(j)$ and
$c_{ij}=|c(i,j)|$ is the number of their common friends. We define
$U=\cup_{q\in n(i)}{n(q)} \cup n(i).$ We also define $\Psi=\{j\} \cup c(i,j),$ where $j$ is a random individual appearing in $i$'s information sequence $s(i)$ and $j\notin n(i)$. Based on the definition of
information entropy in ~\cite{max_entropy}, we can obtain the
information entropy for node $i$ is
\begin{align}
\epsilon(i)_j=&-\sum_{q\in U/\Psi}{\frac{n_q}{s_i}\log{\frac{n_q}{s_i}}} -
\sum_{l\in c(i,j)}{\frac{n_l}{s_i}\log{\frac{n_l}{s_i}}} \notag \\
&-\frac{c_{ij}}{s_i}\log{\frac{c_{ij}}{s_i}},
\end{align}
where $n_q$ is the count that $q$ appears in $s(i)$ and $s_i$ is the
length of $s(i)$. Since we mainly investigate the evolution in local structure,
here only friends of $i$ and friends of its friends are considered during the
computation of the entropy. Then we assume that a new friendship is
established between $i$ and $j$ and the current entropy for $i$ is
\begin{align}
\epsilon{'}(i)_j=&-\sum_{q\in
U/\Psi}{\frac{n_q}{s_i'}\log{\frac{n_q}{s_i'}}} - \sum_{l\in
c(i,j)}{\frac{n_l+1}{s_i'}\log{\frac{n_l+1}{s_i'}}} \notag \\
&-\frac{c_{ij}+1}{s_i'}\log{\frac{c_{ij}+1}{s_i'}}-
(k_j-c_{ij})\frac{1}{s_i'}\log{\frac{1}{s_i'}},
\end{align}
where $s_i{'}=s_i+k_j-c_{ij}+1+c_{ij}=s_i+k_j+1$, which is the
length of the updated information sequence, where $k_j$ is the
initial degree of $j.$ Therefore, the change of entropy for $i$ caused by the new tie with $j,$ i.e., $\Delta \epsilon(i)_j=\epsilon{'}(i)_j-\epsilon(i)_j$ could be
rewritten as
\begin{align}
\Delta \epsilon(i)_j = &\sum_{q\in
U/\Psi}{(\frac{n_q}{s_i}\log{\frac{n_q}{s_i}} -
\frac{n_q}{s_i'}\log{\frac{n_q}{s_i'}})} \notag \\
&+ \sum_{l\in c(i,j)}{(\frac{n_l}{s_i}\log{\frac{n_l}{s_i}}-\frac{n_l+1}{s_i'}\log{\frac{n_l+1}{s_i'}})}\notag \\
&+ (\frac{c_{ij}}{s_i}\log{\frac{c_{ij}}{s_i}} -
\frac{c_{ij}+1}{s_i'}\log{\frac{c_{ij}+1}{s_i'}}) \notag \\
&- (k_j-c_{ij})\frac{1}{s_i'}\log{\frac{1}{s_i'}}.
\end{align}
Assume $f(x) = x\log x$, $$f(x+\Delta x) = f(x) + f'(x)\Delta x +
o({(\Delta x)^2}),$$
therefore,
$$ \frac{n_l+1}{s_i'}\log{\frac{n_l+1}{s_i'}} =
\frac{n_l}{s_i'}\log{\frac{n_l}{s_i'}} + (\log{\frac{n_l}{s_i'}} +
1)\frac{1}{s_i'} + o(\frac{1}{{s_i'}^2})$$
and 
$$\frac{c_{ij}+1}{s_i'}\log{\frac{c_{ij}+1}{s_i'}} =
\frac{c_{ij}}{s_i'}\log{\frac{c_{ij}}{s_i'}} +
(\log{\frac{c_{ij}}{s_i'}}+1)\frac{1}{s_i'} + o(\frac{1}{{s_i'}^2}).$$
Then for Equation (3) we have (for details, see {\em Appendix}),
\begin{align}
\Delta \epsilon(i)_j =& -\frac{k_j+1}{s_i'}\epsilon(i)_j - \sum_{l\in
\Psi}{\frac{1}{s_i'}\log{n_l}} - \frac{c_{ij}+1}{s_i'} \notag \\
&+\frac{k_j+1}{s_i'}\log{s_i'} - (c_{ij} +
1)o(\frac{1}{{s_i'}^2}).
\end{align}
Suppose that $k_j$ is fixed, it can be easily obtained
that as $c_{ij}$ grows, $\Delta \epsilon(i)_j$ decreases. Given the network is
undirected, so this conclusion is also proper for $j$. Then we can
conclude that if we build a new tie between $i$ and $j$, the
information entropy gain $\Delta \epsilon(i,j)=\Delta
\epsilon(i)_j+\Delta \epsilon(j)_i$ produced by this new friendship for
the two nodes decreases as $c_{ij}$ increases. It tells us that for
the nodes with more common friends, establishing a new tie between
them produces less information entropy gain for them. Be brief, there is a competition between homophily and information entropy in breeding a new connection. Note that $\Delta \epsilon(i,j)$ declining with $c_{ij}$ might be very slow, because generally $s_i'$ is much greater than $c_{ij}.$

In fact, the information entropy for $i$ represents the diversity of its
information sources. If we create ties between $i$ and other nodes
who have overlapped friends with it, these nodes will appear more
frequently in its information sequence and even become the dominating
sources of the information. Then the diversity of the information
source is weaken and the gain of the information entropy decays accordingly.

\section{Empirical Analysis}
\label{sec:ea}
In order to validate the above analysis, we employ several data
sets, including both synthetic and real-world networks, for further empirical study. The
synthetic data sets are generated by BA~\cite{ba_model}, Small
World~\cite{sw_model} and CNNR~\cite{cnnr} models. BA is a classic
model to generate scale-free networks with the mechanism of
preferential attachment. We denote the data set it generates as
BA$(N,m)$, where $N$ is the size of the network and $m$ is the number of initial ties that would be connected when a new node is
added. Small World model is a random model with probability $p$ to
rewire and produce long range ties, it can be denoted as
SW$(N,K,p)$, where $2K$ is the averaged degree. CNNR model is
modified from CNN~\cite{cnn} for generating social networks,
especially online social networks. We denote it as CNNR$(N,u,r)$,
where $u(1-r)$ is the probability to covert the potential edges into
real ties. The averaged degree of the network it generates is
approximately $2/(1-u)$. The real-world data sets come from
different fields. For example, \texttt{CA-HepPh} is a collaboration
network from the e-print arXiv\footnote{http://www.arxiv.org} and covers scientific collaborations
between authors of papers submitted to High Energy
Physics~\cite{ca_hepPh_dataset}. \texttt{NewOrleans} is the Facebook
network in New Orleans~\cite{neworleans_dataset}.
\texttt{Email-Enron} is an email communication network that covers
all the email communication within a data set of around half million
emails~\cite{email_dataset}. The basic properties of theses data
sets we utilize in following experiments are listed in
Tab.~\ref{tab:dataset}.

\begin{table}
\caption{Data Sets\label{tab:dataset}}
\begin{center}
\begin{tabular}{lll}
\hline\noalign{\smallskip}
Data set & $N$ & $|E|$\\
\noalign{\smallskip}\hline\noalign{\smallskip}
\texttt{BA(20000,10)} & 20000 & 199352 \\
\texttt{SW(20000,10,0.1)} & 20000 & 200000 \\
\texttt{CNNR(20000,0.9,0.04)} & 20000 & 187215\\
\texttt{CA-HepPh} & 12006 & 118489 \\
\texttt{NewOrleans} & 63392 & 816886 \\
\texttt{Email-Enron} & 36692 & 183831\\
\noalign{\smallskip}\hline
\end{tabular}
\end{center}
\end{table}

\begin{figure}
\subfloat[\texttt{BA(20000,10)}]{\includegraphics[width=1.7in]{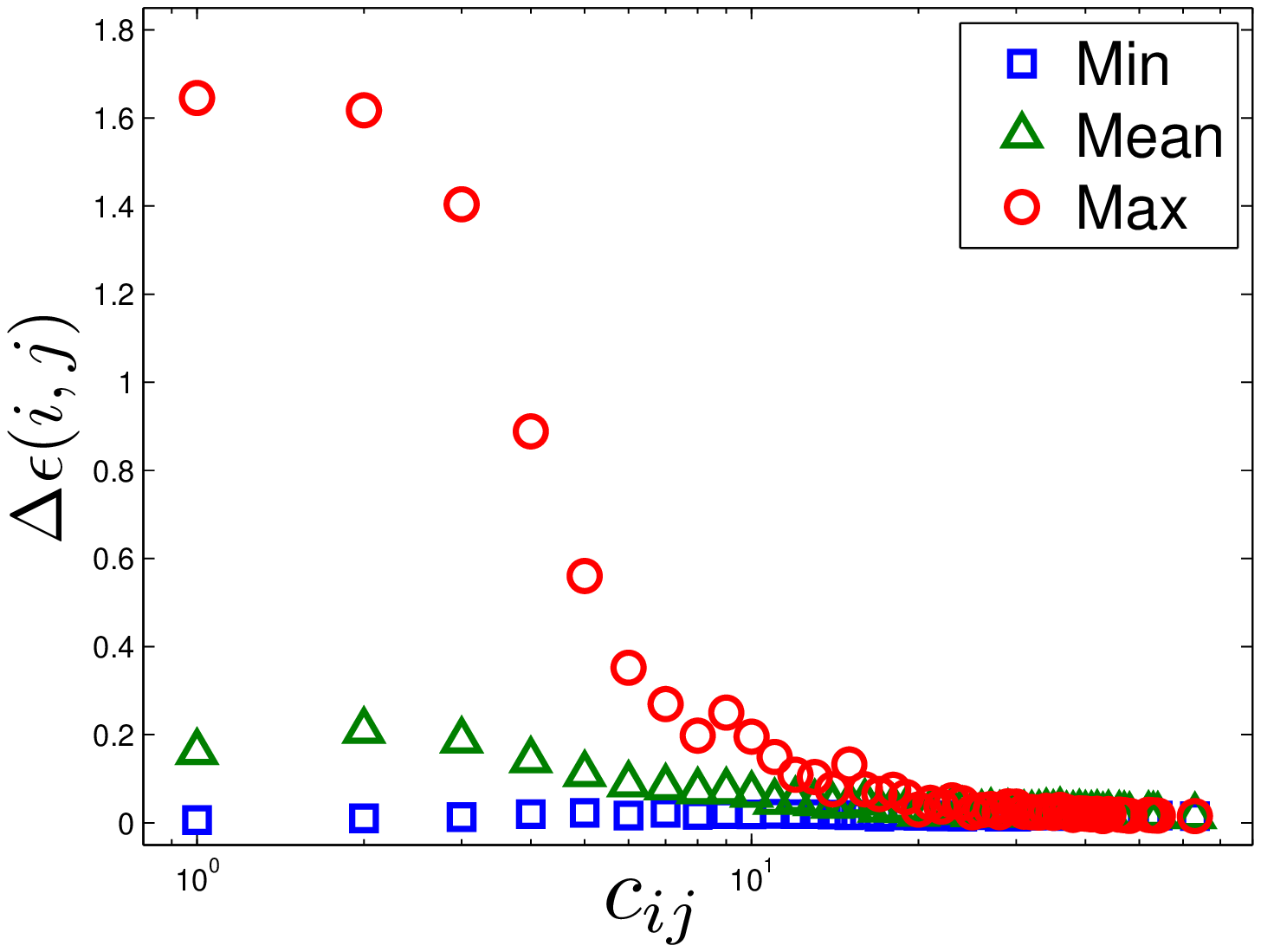}
 \label{fig:ba_cnn_delta_e}}
\subfloat[\texttt{SW(20000,10,0.1)}]{\includegraphics[width=1.7in]{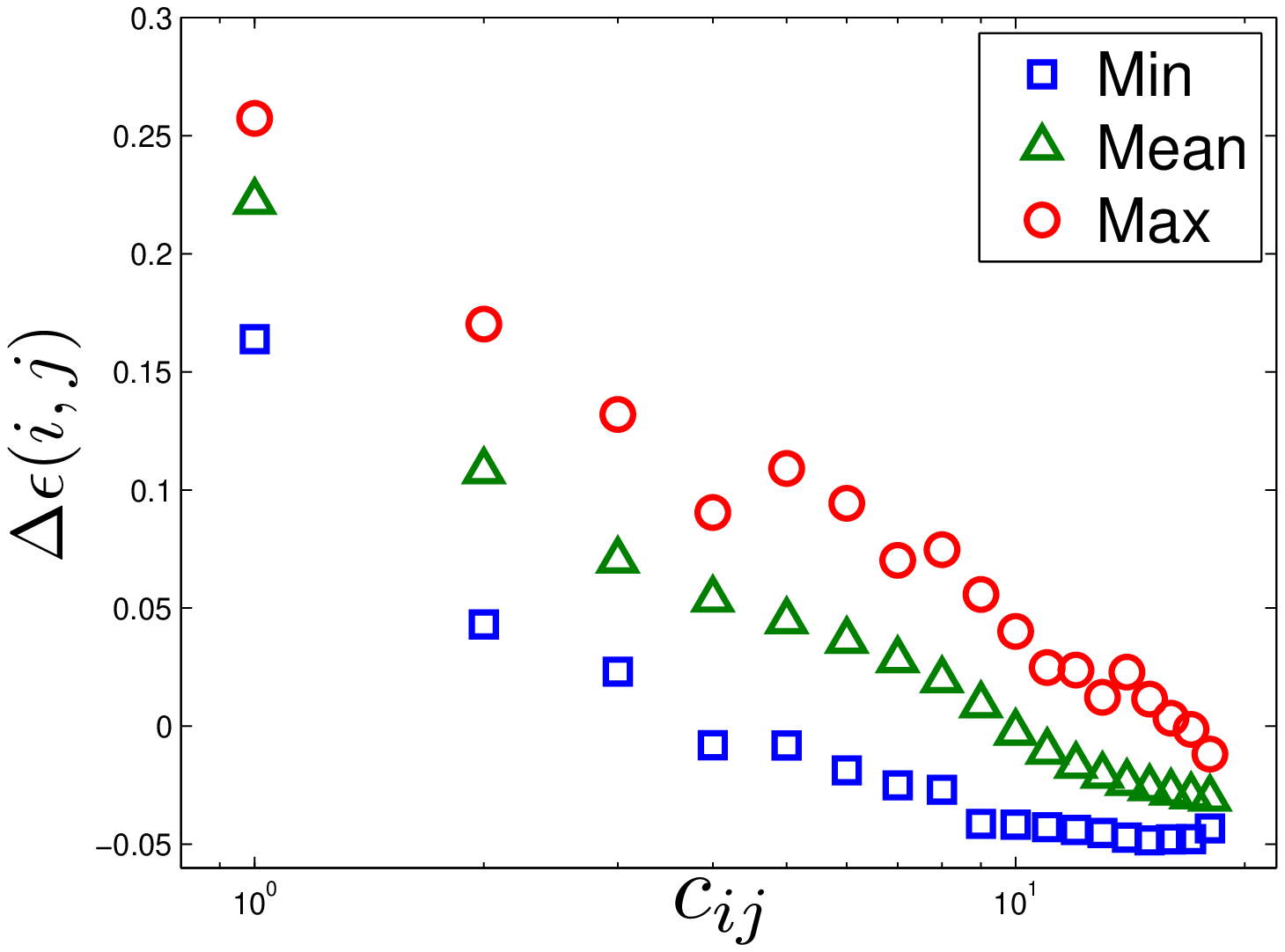}
 \label{fig:sw_cnn_delta_e}}\\
\subfloat[\texttt{CNNR(20000,0.9,0.04)}]{\includegraphics[width=1.7in]{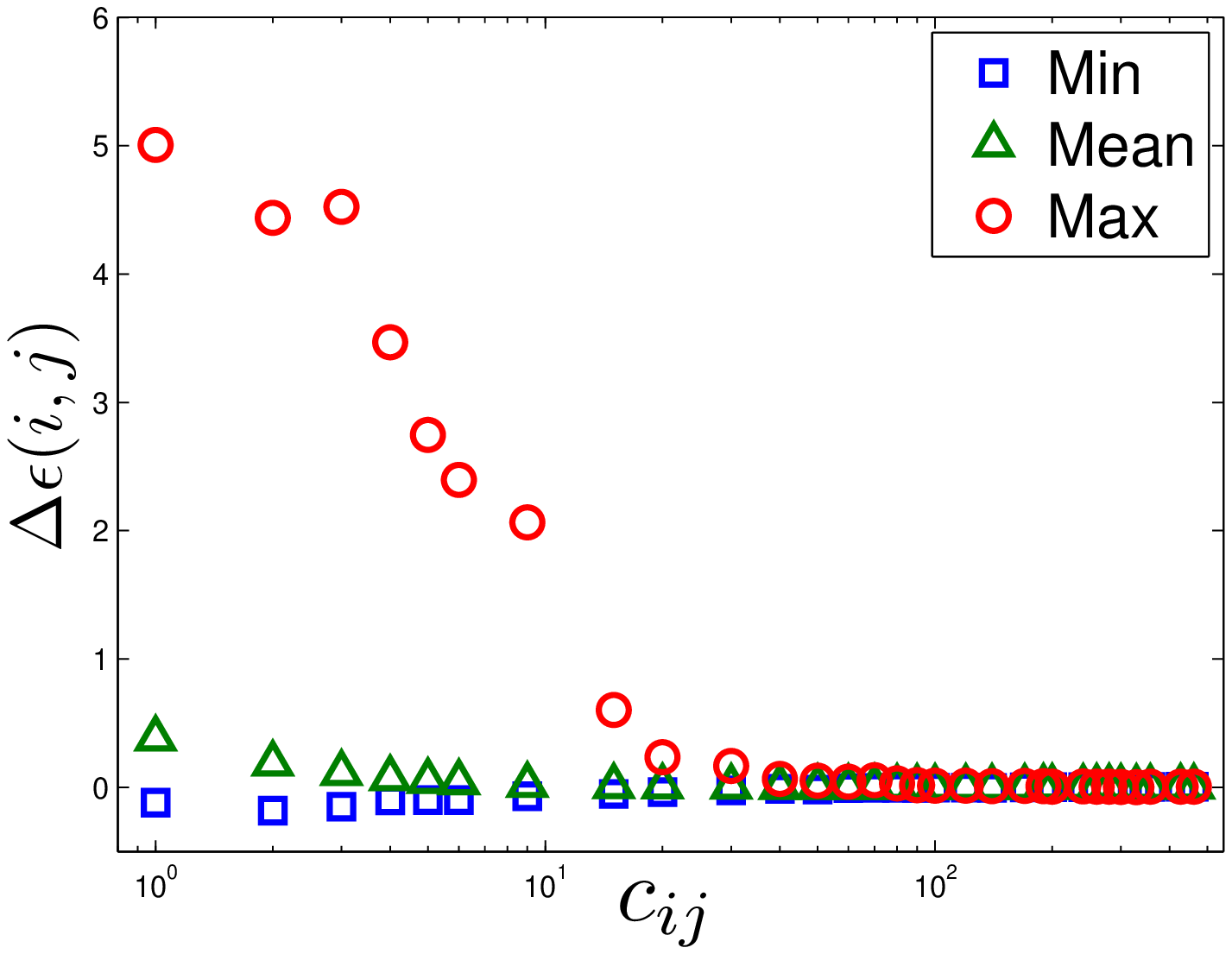}
 \label{fig:cnnr_cnn_delta_e}}
\subfloat[\texttt{NewOrleans}]{\includegraphics[width=1.7in]{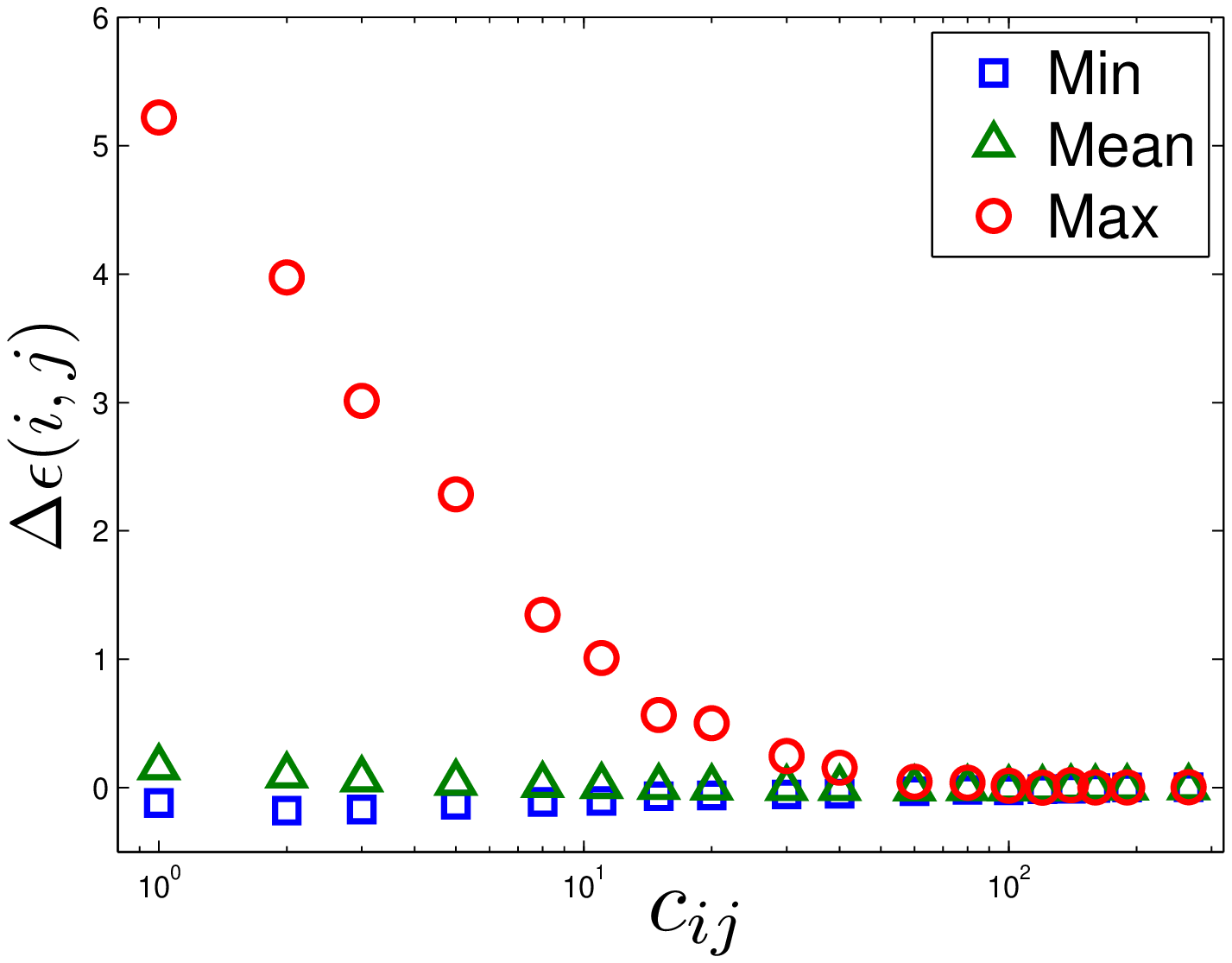}
 \label{fig:neworleans_cnn_delta_e}}\\
\subfloat[\texttt{CA-HepPh}]{\includegraphics[width=1.7in]{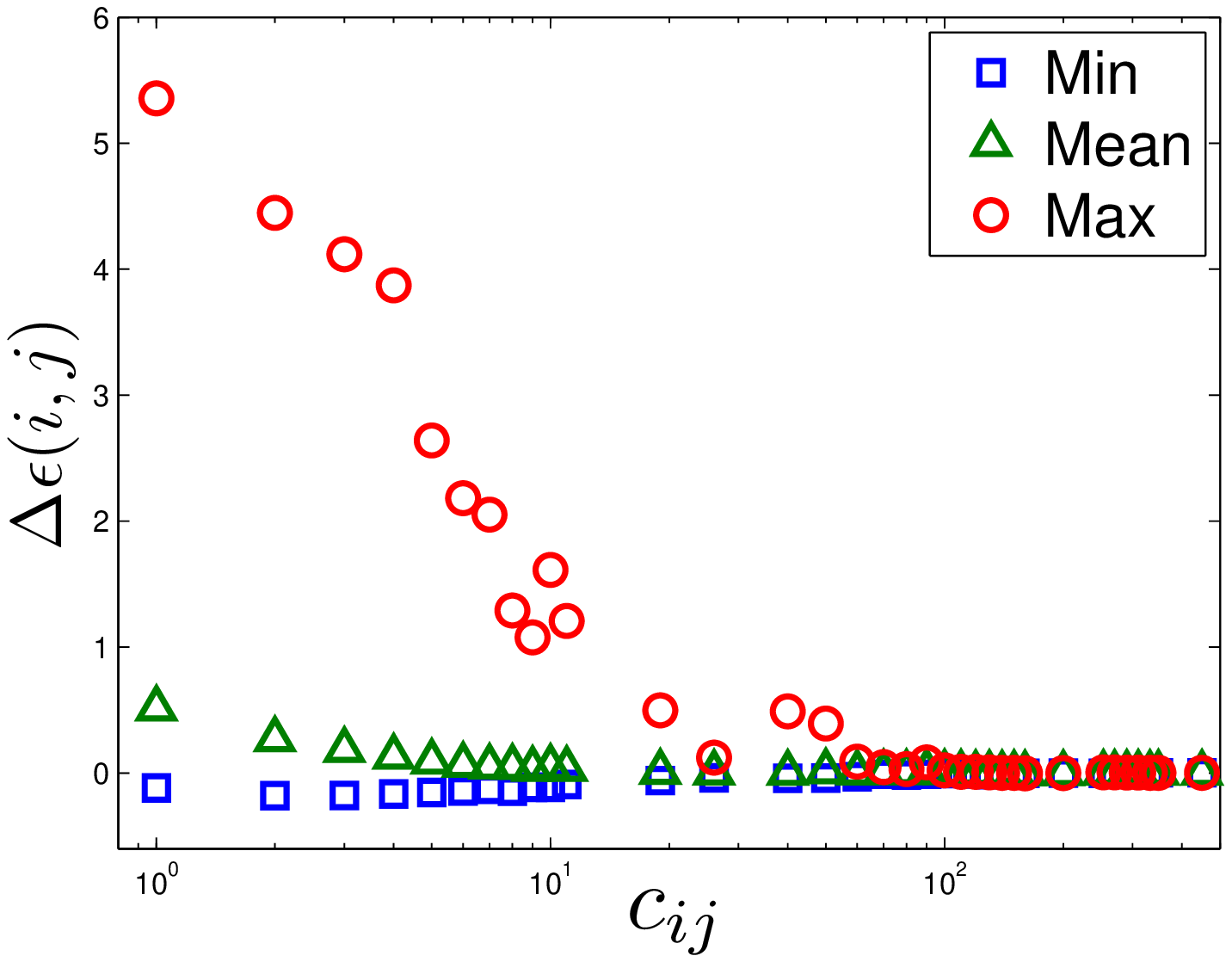}
 \label{fig:cahepph_cnn_delta_e}}
\subfloat[\texttt{Email-Enron}]{\includegraphics[width=1.7in]{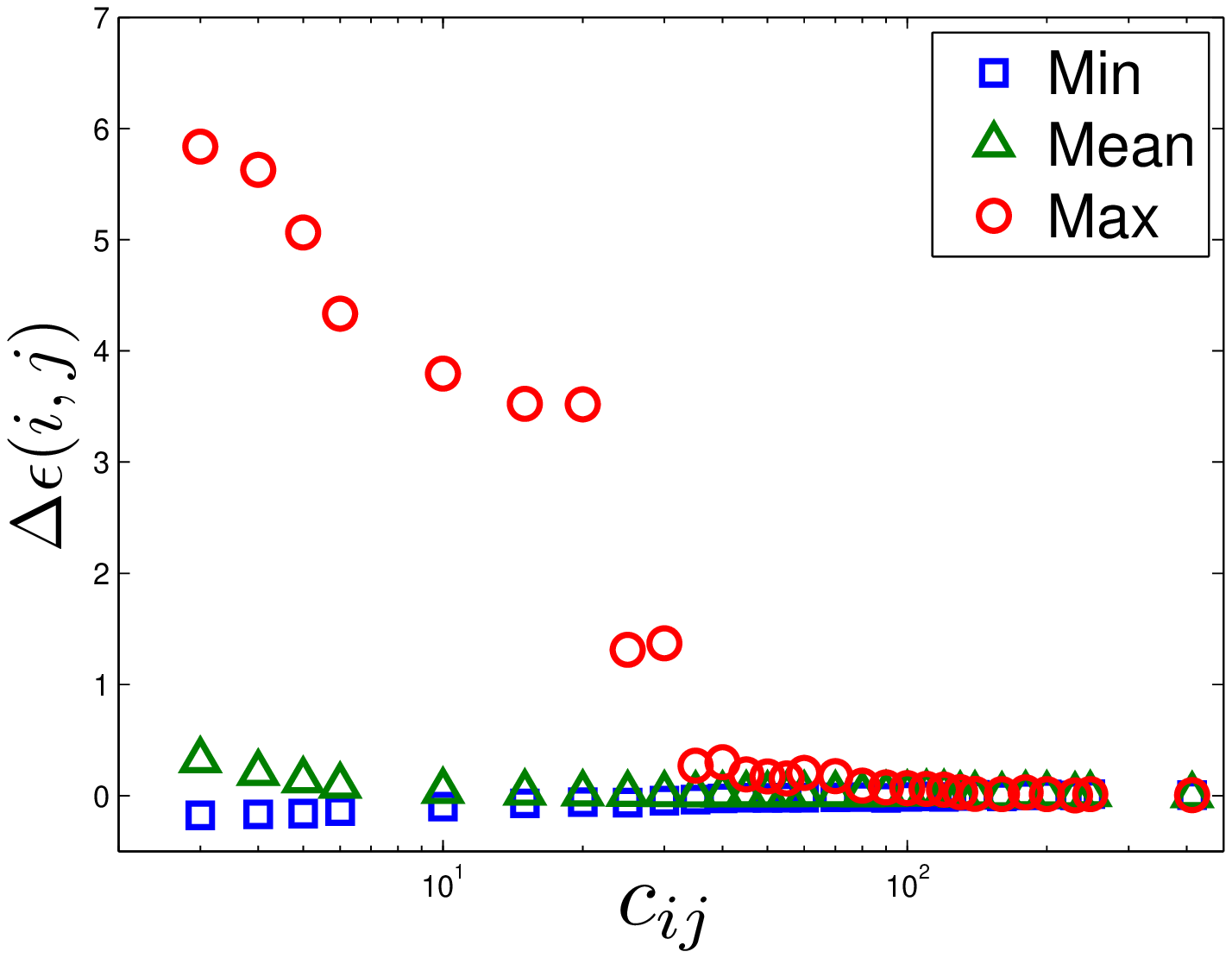}
 \label{fig:email_cnn_delta_e}}
\caption{Empirical results from data sets. The results are consistent with the theory that increment of common friends would decrease the information entropy gain, especially for the maximum. Particularly, it should be also noted that as predicted by the analytical results, the averaged decay of $\Delta \epsilon(i,j)$ is very small in some cases, as shown in FIG.~\ref{fig:neworleans_cnn_delta_e}. Note that there are several outliers for the maximum $\Delta \epsilon(i,j)$, like in FIG.~\ref{fig:cnnr_cnn_delta_e}, which are produced by the noise in statistics. While the global trend of decrement with $c_{ij}$ in all networks is still significant.}
\label{fig:cnn_delta_e}
\end{figure}

As discussed before, establishing a new friendship may affect the
entropy of the both ends. In the above networks, we characterize the
relation between $c_{ij}$ and $\Delta \epsilon(i,j)$ in the
following steps: For each tie between $i$ and $j$, we first obtain
$\epsilon{'}(i)_j+\epsilon{'}(j)_i$ in the origin network; Secondly,
we delete this tie and get $\epsilon(i)_j+\epsilon(j)_i$; Thirdly,
the tie is restored. For different $\Delta \epsilon(i,j)$ for the
same $c_{ij}$, we get the maximum, mean and minimum values,
respectively. The change of entropy for other nodes in the
network is not considered here for the reason that we assume the
establishment of a tie between $i$ and $j$ is a personal activity
with local information solely. As shown in
FIG.~\ref{fig:cnn_delta_e}, in all networks, $\Delta
\epsilon(i,j)$ decreases as $c_{ij}$ grows, which is consistent with
our above analysis, especially for the small world network in FIG.~\ref{fig:sw_cnn_delta_e}. At the start stage, the diverge between the
maximum and mean of $\Delta \epsilon(i,j)$ is large, then it decays
quickly as $c_{ij}$ increases. It is also observed that for the
nodes with tremendous common friends, building a new friendship
between them may even lead to entropy loss. To sum up, the empirical results testify our statement further that increment of homophily would reduce the information entropy gain, which indicates a competition between the two evolving rules. 

\section{Discussion}
\label{sec:discuss}
\begin{table}[h!]
\caption{$\tau$ of the real-world networks.\label{tab:real_c_e}}
\begin{center}
\begin{tabular}{lll}
\hline\noalign{\smallskip}
Data set & $\tau$ & $c$ \\
\noalign{\smallskip}\hline\noalign{\smallskip}
\texttt{NewOrleans} & 0.70 & 0.22\\
\texttt{Email-Enron} & 0.56 & 0.50 \\
\texttt{CA-HepPh} & 0.50 & 0.61 \\
\noalign{\smallskip}\hline
\end{tabular}
\end{center}
\end{table}

\begin{figure}
\centering
\includegraphics[width=3in]{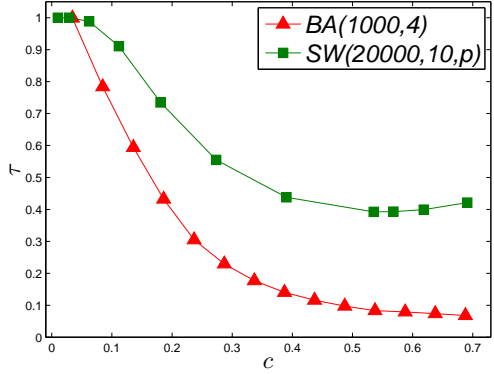}%
\caption{$\tau$ of the network varies as $c$ increases.
\label{fig:c_e}}
\end{figure}

The growing of a social network could be simply regarded as establishing
new ties among individuals. From the perspective of information
entropy maximization, a tie should be established to gain more
entropy for both ends. Therefore, we could distinguish the tie that
makes the entropy of its ends gain as the positive tie, while the
one that leads to entropy loss as the negative tie. Then we define
the positiveness of the social network as the fraction of positive
ties, which is denoted as $\tau$. Larger $\tau$ means more ties in
the network are established to increase their ends' entropy gain.  As
shown in Tab.~\ref{tab:real_c_e}, we list $\tau$ of the real-world
network, where $c$ is the clustering of the network. It is
interesting that for the network with higher $c$, its
$\tau$ is lower generally. We also investigate this finding on the
network with various clusterings generated by BA and Small World
models. For the BA model, we employ the method of tuning clustering
while keeping its degree distribution stable~\cite{kim_clustering,ma_clustering}. We only perform experiments of tuning the clustering
on BA(1000,4), because it is too much time consuming for BA(20000,10). For
the model of Small World, we just vary $p$. As shown in
FIG.~\ref{fig:c_e}, for both of models, the positiveness of
network decreases as $c$ grows. In fact, the clustering of the
network could be rewritten~\cite{rewrite_c} as
$$c=\frac{1}{|V|}\sum_{\forall (i,j) \in E}{\frac{c_{ij}}{{k_i \choose 2}}}.$$
For this reason, with respect to the rule of homophily, a new tie
added preferentially between nodes with overlapped friends would
also lead to new triangles constructed in local structure. That is
to say, the clustering of the network, i.e., $c,$ would be increased when its
evolution is driven by the homophily. Because of this, homophily
dominated evolution leads to the decrement of $\tau$. However, with respect to the information entropy maximization, the new tie is
established to increase the diversity of the information source and
gain more entropy, which would improve $\tau$ by importing more positive ties.

\begin{figure}
\centering
\subfloat[\texttt{BA(1000,4)}]{\includegraphics[width=1.7in]{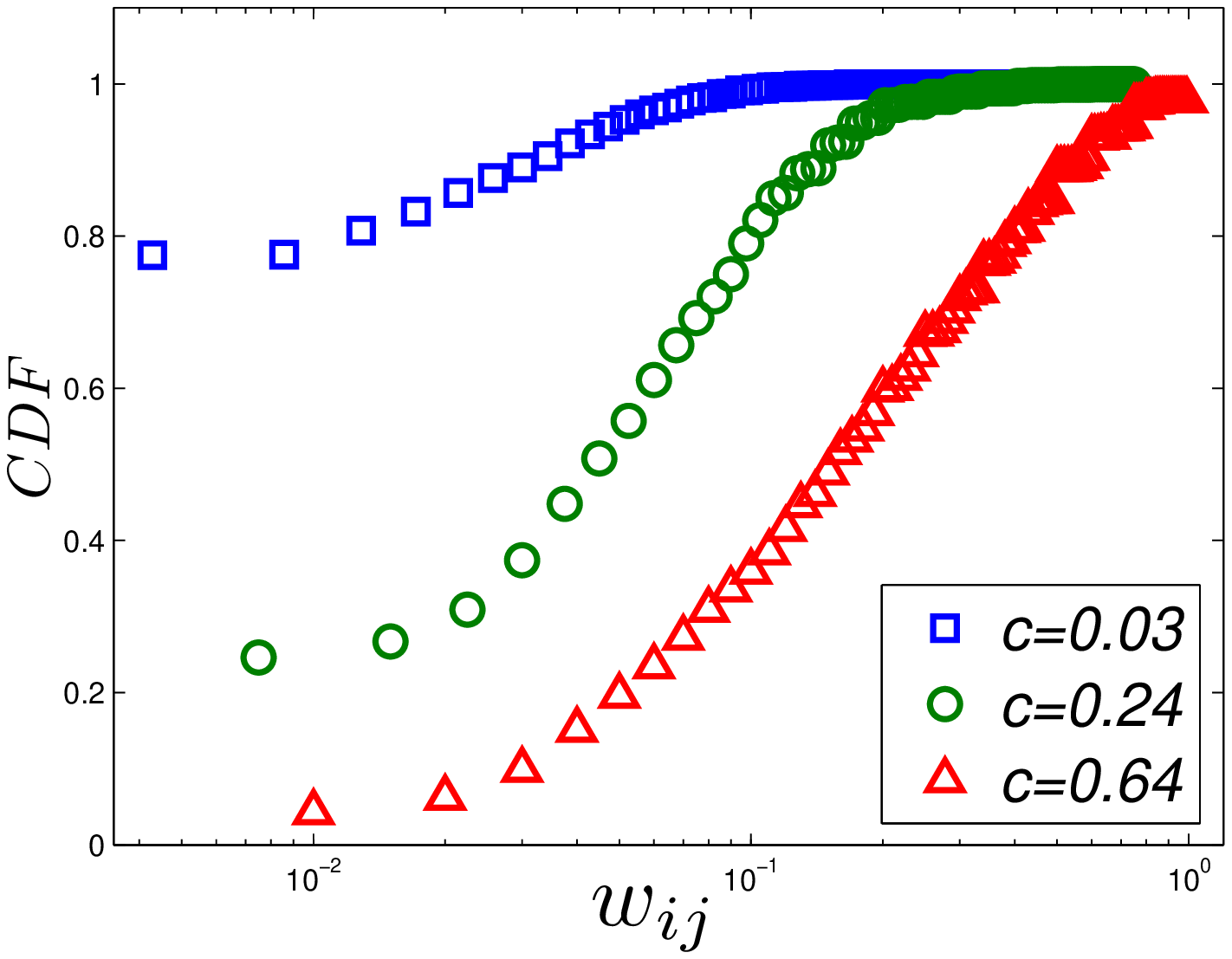}
 \label{fig:ba_c_cdf}}
\subfloat[\texttt{SW(20000,10,p)}]{\includegraphics[width=1.6in]{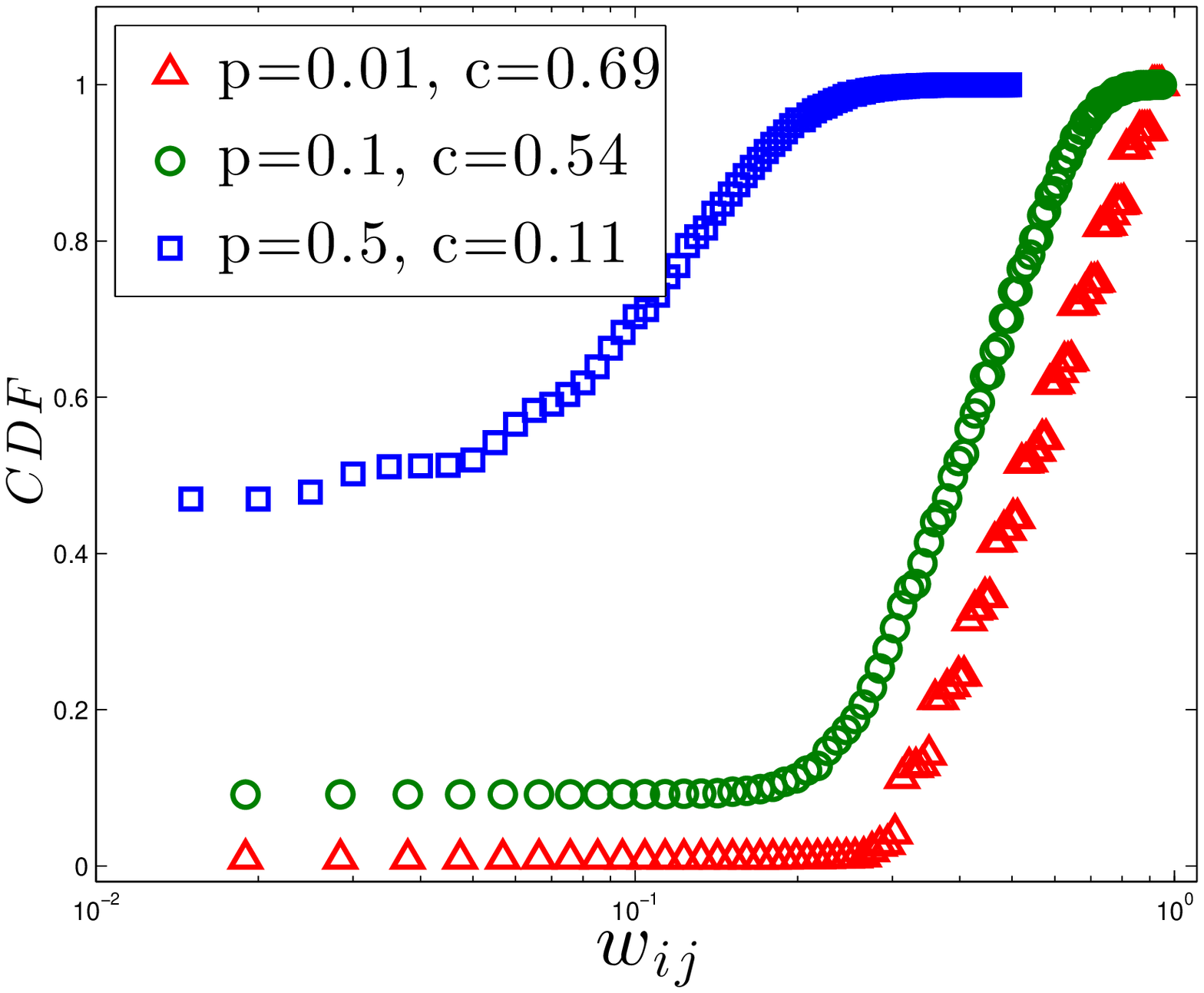}
 \label{fig:sw_c_cdf}}
\subfloat[\texttt{Real-world
networks}]{\includegraphics[width=1.7in]{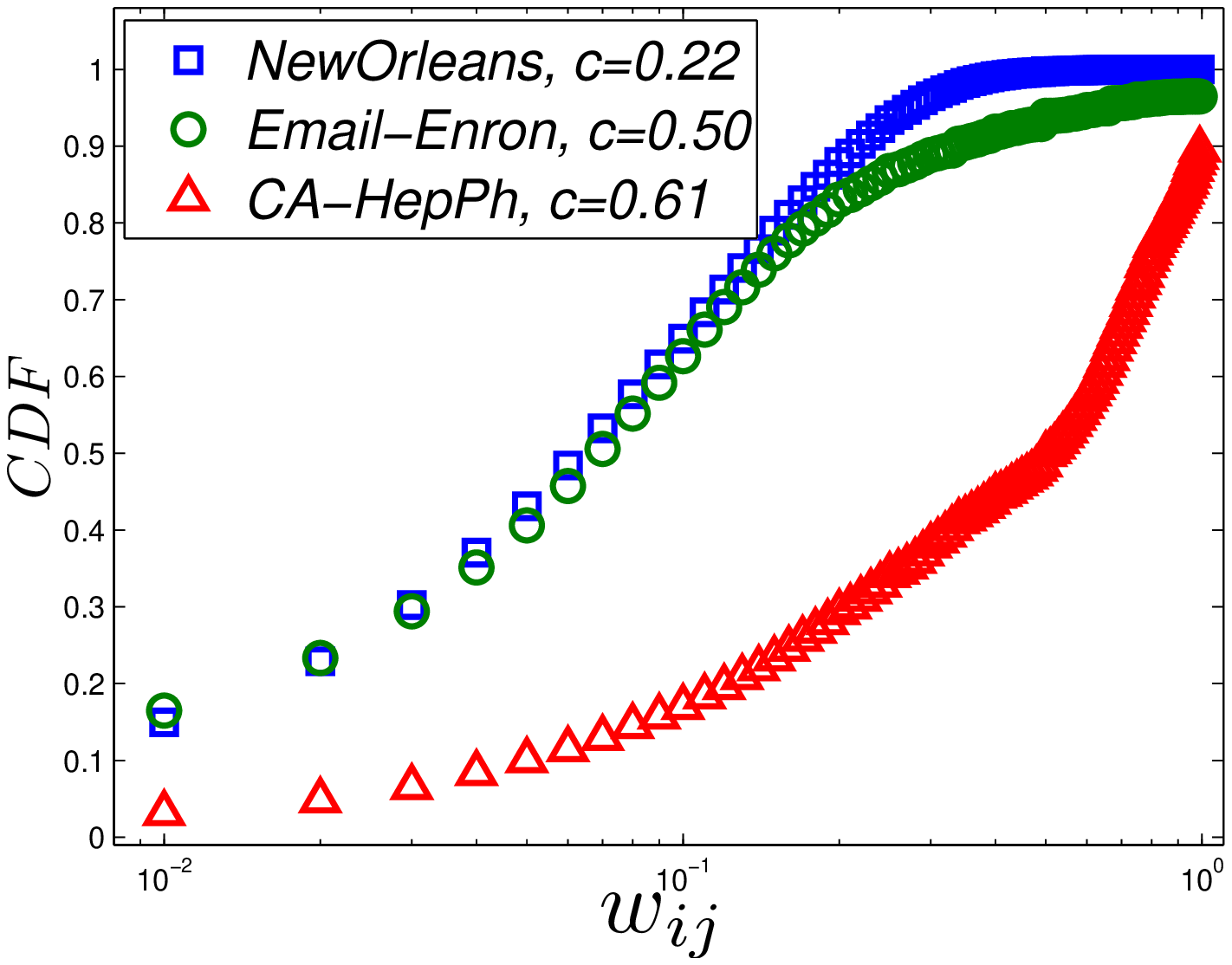}
 \label{ig:realworld_c_cdf}}
\caption{CDF of $w_{ij}$ for various $c$.}
\label{fig:c_cdf}
\end{figure}

The strength of a social tie can be defined as the number of
overlapped friends between its ends. For example, the strength of a
tie between $i$ and $j$ could be defined as
$w_{ij}=c_{ij}/(k_i-1+k_j-1-c_{ij})$~\cite{mobile_network,bridgeness,tie-role}, where lower $w_{ij}$ stands for a weak tie. It is obvious
that if $i$ and $j$ share a lot of common friends, the strength of
the tie between them is strong. Conventionally, it is thought that
the weak tie is helpful in getting the new
information~\cite{weakties}, while the strong tie means the
relationship is trustful~\cite{trustties}. Therefore, based on the
above discussion, it seems that the evolution supervised by
homophily could lead to generations of strong ties in the network,
because it renders the network highly clustered. In order to
validate this, we observe the cumulative distribution function(CDF)
of $w_{ij}$ for each tie in the network. As shown in
FIG.~\ref{fig:c_cdf}, as $c$ of the network decreases, the CDF
curve moves to the left, which indicates the increment of the
fraction of weak ties~\cite{cdf_wij}. It validates our conjecture
that in both synthetic and real-world data sets, highly clustered
networks caused by homophily contain more strong ties, while the
ones with lower clusterings contain more weak ties, which are produced by the law of maximum information entropy.

\section{Conclusion and Future Work}
\label{sec:conclusion}
In summary, both theoretical analysis and experimental results show
that the rule of homophily is competing with the law of
information entropy maximization in social networks. Moreover, the rule of homophily
driven evolution makes the network highly clustered and increases
the certainty of the information source for a node. Contrarily, the rule of maximum entropy leads to the diversity of information sources. Based on the definition of
weak ties, we can conclude that the rule of maximum information entropy leads
to the generation of weak ties in the network, while the homophily
produces strong ties between nodes with overlapped friends.
Corresponding to the fact that both the weak and strong ties coexist
in the network, we conjecture that both of the evolving rules might
coexist in growth of the social networks. Therefore, in the view of
maximum information entropy, the social network is not efficient, however, it
owns many strong ties which may deliver trust information. Our findings could provide insights for modeling social network evolution as a competition of different rules.

Given the tremendous development of the online social network, the cost of social activity in the epoch of the Internet continues to decrease~\cite{trustties,ahn-cyworld-wwww}. Because of this, we neglect the cost of establishing ties of different strengths for simplifying the analytical framework in this paper. While in the real world, the social activity is constrained by the personal cognition limit and social cost~\cite{offlinenetworksize} and the Dunbar's number~\cite{dunbarnumber} still exists in the online social network~\cite{dunbar-zhao,golder:rhythms,ahn-cyworld-wwww}. Hence in the future work, we would take the cost of establish different ties into consideration and build an evolution model of social networks based on the competition of strong and weak ties. 

\section*{Acknowledgment}
\label{sec:ack}
Jichang Zhao was partially supported by the Fundamental Research Funds for the Central Universities (Grant Nos. YWF-14-RSC-109 and YWF-14-JGXY-001).


\section*{Appendix}
\label{sec:append}
\allowdisplaybreaks
\begin{eqnarray*}
\Delta \epsilon(i)_j &=& \sum_{q\in
U/\Psi}{(\frac{n_q}{s_i}\log{\frac{n_q}{s_i}} -
\frac{n_q}{s_i'}\log{\frac{n_q}{s_i'}})} \\ &+& \sum_{l\in
c(i,j)}(\frac{n_l}{s_i}\log{\frac{n_l}{s_i}} -
\frac{n_l}{s_i'}\log{\frac{n_l}{s_i'}} - (\log{\frac{n_l}{s_i'}} +
1)\frac{1}{s_i'}\\&-&o(\frac{1}{{s_i'}^2})) \\
&+& (\frac{c_{ij}}{s_i}\log{\frac{c_{ij}}{s_i}} -
\frac{c_{ij}}{s_i'}\log{\frac{c_{ij}}{s_i'}} - (\log{\frac{c_{ij}}{s_i'}} +
1)\frac{1}{s_i'}-o(\frac{1}{{s_i'}^2})) \\
&-& (k_j - c_{ij})\frac{1}{s_i'}\log{\frac{1}{s_i'}} \\
&=& \sum_{q\in U}{(\frac{n_q}{s_i}\log{\frac{n_q}{s_i}} -
\frac{n_q}{s_i'}\log{\frac{n_q}{s_i'}})} - \sum_{l\in
\Psi}{\frac{1}{s_i'}(\log{\frac{n_l}{s_i'}} + 1)}\\
&-& (k_j-c_{ij})\frac{1}{s_i'}\log{\frac{1}{s_i'}} - (c_{ij} +
1)o(\frac{1}{{s_i'}^2}) \\
&=& \sum_{q\in U}{(\frac{n_q}{s_i}\log{\frac{n_q}{s_i}} -
\frac{n_q}{s_i'}(\log{\frac{n_q}{s_i}} + \log{\frac{s_i}{s_i'}}))}\\ &-&
\sum_{l\in \Psi}{\frac{1}{s_i'}(\log{\frac{n_l}{s_i'}} + 1)} -
(k_j-c_{ij})\frac{1}{s_i'}\log{\frac{1}{s_i'}}\\ &-& (c_{ij} +
1)o(\frac{1}{{s_i'}^2}) \\
&=& (1-\frac{s_i}{s_i'})\sum_{q\in U}{\frac{n_q}{s_i}\log{\frac{n_q}{s_i}}} - \frac{s_i}{s_i'}\log{\frac{s_i}{s_i'}}\\
&-& \sum_{l\in \Psi}{\frac{1}{s_i'}(\log{\frac{n_l}{s_i'}} + 1)} -
(k_j-c_{ij})\frac{1}{s_i'}\log{\frac{1}{s_i'}} \\&-& (c_{ij} +
1)o(\frac{1}{{s_i'}^2})\\
&=& -(1-\frac{s_i}{s_i'})\epsilon(i)_j - \sum_{l\in
\Psi}{\frac{1}{s_i'}(\log{\frac{n_l}{s_i'}} + 1)}\\& -&
(k_j-c_{ij})\frac{1}{s_i'}\log{\frac{1}{s_i'}} - (c_{ij} +
1)o(\frac{1}{{s_i'}^2}) \\
&=& -(1-\frac{s_i}{s_i'})\epsilon(i)_j - \sum_{l\in
\Psi}{\frac{1}{s_i'}\log{\frac{n_l}{s_i'}}} - \frac{c_{ij}+1}{s_i'} \\&-&
(k_j-c_{ij})\frac{1}{s_i'}\log{\frac{1}{s_i'}} - (c_{ij} +
1)o(\frac{1}{{s_i'}^2})\\
&=& -(1-\frac{s_i}{s_i'})\epsilon(i)_j - \sum_{l\in
\Psi}{\frac{1}{s_i'}\log{n_l}} - \frac{c_{ij}+1}{s_i'} \\&+&
(k_j+1)\frac{1}{s_i'}\log{s_i'} - (c_{ij} +
1)o(\frac{1}{{s_i'}^2})\\
&=& -\frac{k_j+1}{s_i'}\epsilon(i)_j - \sum_{l\in
\Psi}{\frac{1}{s_i'}\log{n_l}} - \frac{c_{ij}+1}{s_i'}\\ &+&
\frac{k_j+1}{s_i'}\log{s_i'} - (c_{ij} +
1)o(\frac{1}{{s_i'}^2})\\
\end{eqnarray*}

\end{document}